\documentclass{gCOM2e}

\usepackage{epstopdf}
\usepackage{subfigure}
\usepackage{multirow}
\usepackage{xr}
\usepackage{afterpage}
\usepackage{hhline}
\usepackage{float}

\theoremstyle{plain}

\theoremstyle{definition}

\theoremstyle{remark}

\begin{document}


\title{\textit{Speed and biases of Fourier-based pricing choices: A numerical analysis}}

\author{
\name{Ricardo Cris\'ostomo\textsuperscript{a}$^{\ast}$\thanks{$^\ast$Corresponding author. Email: rcayala@cnmv.es}}
\affil{\textsuperscript{a}Comisi\'on Nacional del Mercado de Valores (CNMV) and Department of Statistics and Operations Research, UNED}}

\maketitle

\begin{abstract}
We compare the CPU effort and pricing biases of seven Fourier-based implementations. Our analyses show that truncation and discretization errors significantly increase as we move away from the Black-Scholes-Merton framework. We rank the speed and accuracy of the competing choices, showing which methods require smaller truncation ranges and which are the most efficient in terms of sampling densities. While all implementations converge well in the Bates jump-diffusion model, Attari's formula is the only Fourier-based method that does not blow up for any Variance Gamma parameter values. In terms of speed, the use of strike vector computations significantly improves the computational burden, rendering both fast Fourier transforms (FFT) and plain delta-probability decompositions inefficient. We conclude that the multi-strike version of the COS method is notably faster than any other implementation, whereas the strike-optimized Carr Madan's formula is simultaneously faster and more accurate than the FFT, thus questioning its use.

\end{abstract}


\begin{classcode}30E10; 60H35; 65C20; 65T50
\end{classcode}

\section{Introduction}\label{s:1}

Since the seminal papers of Black-Scholes and Merton \cite{Black1973, Merton1973}, processes where asset prices diffuse continuously have been extensively used in risk management and option pricing. Diffusion models exhibit a variety of forms, including stochastic volatility, mean-reversion or seasonality, and their widespread use highlights the success that these models have achieved in financial modelling. Yet casual observation reveals that the prices of traded assets routinely undergo jumps. Discontinuities can occur, for instance, due to unexpected news, due to trading restrictions or simply because there is a substantial imbalance between buy and sell orders.

The importance of jump modelling becomes evident if we analyze the prices of short dated out-of-the-money (OTM) options. The value of these contracts critically stems from an expectation of large underlying movements. However, empirical studies have shown that diffusion-only models cannot consistently generate the asymmetry and fat-tails that are routinely implied by short-term OTM options \cite{Bakshi1997a, Cont2004}.

This paper contributes to the option pricing literature by benchmarking the speed and accuracy of seven Fourier-based pricing choices. Specifically, our analyses focus on two jump models that have been proposed as a framework to price options with different strikes and maturities. First, the Bates jump-diffusion model \cite{Bates1996a}, which blends the Heston dynamics with lognormally distributed price jumps. Second, the Asymmetric Variance Gamma (AVG) \cite{Madan1998a}, a purely discontinuous process where the underlying assets evolve through a combination of many small jumps and rare big moves.

Both models are implemented by means of characteristic functions. Fourier transforms are rapidly gaining traction in finance and most of the option pricing models developed in the last decade have relied on characteristic functions to obtain option prices. Thus, a better understanding of the different implementations is paramount to avoid pricing errors. We investigate the speed and biases of a wide range of Fourier pricing choices, including Delta-probability decompositions, the Carr-Madan and Attari formulae, the COS method, and fast Fourier transforms.

The novelty of our paper lies in:
\begin{enumerate}
  \item[\textbf{1.}] We are the first to consider the strike-optimized version of the Carr-Madan and Attari formulas, and one of the first to benchmark the multi-strike version of the COS method. We show that all these alternatives significantly outperform the FFT.
  \item[\textbf{2.}] We compare the numerical efficiency of seven Fourier-based alternatives, showing which methods require the highest/lowest integration range and the highest/lowest sampling densities.
  \item[\textbf{3.}] We find that Attari's formula is the only method that does not blow up in any problematic region of the AVG model.
  \item[\textbf{4.}] We show that the strike-optimized version of Carr-Madan's formula is simultaneously faster and more accurate than the FFT, questioning its widespread use.
\end{enumerate}

An important reference in this respect is the BENCHOP competition \cite{VonSydow2015}. This project compares the accuracy and speed of several Fourier methods, finding that the COS formula is the overall fastest alternative. To benchmark our results to this project, we employ the BENCHOP implementation for the COS method developed by Ruijter and Oosterlee \cite{Ruijter2015}, which we have adapted to simultaniously calculate option prices for different strikes.

The rest of the paper is organized as follows. Section \ref{s:2} reviews the use of characteristic functions and explains the numerical setup. Section \ref{s:3} present the Bates model and compares the accuracy and speed of the different implementations. Section \ref{s:4} describes the AVG model and considers three regions where Fourier methods can lead to notably different accuracies. Finally, section \ref{s:5} summarizes our conclusions.

\section{Characteristic functions for option pricing}\label{s:2}

Under no-arbitrage, option prices can be calculated as the discounted risk-neutral expectation of its terminal payoff

\begin{equation}\label{V0}
  V_0= e^{-rT} ~ \mathbb{E_Q} [ H (S_t) ],
\end{equation}
\noindent where $V_0$ is the option value at time $t  = 0$, $S_t$  the underlying price, $r$  the risk-free rate, $T$ the time to maturity, $H (S_t)$  is the option payoff and $\mathbb{E_Q}[\bullet ]$ denotes the expectation operator under the risk-neutral measure.
For many pricing processes, the expected option payoff can be computed in terms of the underlying asset's density function. For instance, the payoff of a European call with strike $K$ and expiration $T$ is given by $H({S_t}) = {({S_T} - K)^ + }$. Thus, its present value at time $t = 0$ can be obtained as

\begin{equation}\label{C0}
C(T,K) = {e^{ - rT}}\int_0^\infty  {{{({S_T} - K)}^ + }} q({S_T})d{S_T},
\end{equation}

\noindent where $q({S_T})$ is the risk-neutral density of the underlying asset  $S_t$  at the terminal date $T$.

However, there are numerous asset processes that do not exhibit a tractable density. For these cases, pricing models generally rely on characteristic functions in order to obtain option prices. Characteristic functions are defined as the Fourier transform of the probability density functions. Thus, both functions exhibit a one-to-one correspondence and all the probabilistic evaluations that can be performed through a tractable density can be also obtained with characteristic functions. Furthermore, the characteristic functions of many asset specifications, particularly in connection to stochastic volatility and jumps, exhibit simpler and more tractable forms than their corresponding density functions.

\subsection{The Delta-Probability Decomposition (DPD)}\label{s:2.1}

The DPD was initially developed by Heston \cite{Heston1993}. By expanding \eqref{C0}, it is straightforward to show that the price of a European call can be expressed as

\begin{equation}\label{CDPD}
C(T,K) = {S_0}{\rm{ }}{\Pi _1} - {e^{ - rT}}K{\rm{ }}{\Pi _2},
\end{equation}

\noindent where  $\Pi _1$ and  $\Pi _2$  are two probability-related quantities. Specifically,  $\Pi _1$  is the option delta while $\Pi _2$  is the risk-neutral probability of exercise ${\mathop{\rm P}\nolimits} ({S_T} > K)$.

In the Black-Scholes-Merton (BSM) model and other simple processes, these probabilities can be directly computed in terms of the underlying asset density function. However, for processes that do not exhibit a tractable density, Bakshi and Madan \cite{Bakshi2000} show that these probabilities can also be computed as

\begin{equation}\label{pi1}
{\Pi _1} = \frac{1}{2} + \frac{1}{\pi }\int_0^\infty  {{\mathop{\rm Re}\nolimits} \left[ {\frac{{{e^{ - iw\ln (K)}}{\psi _{\ln {S_T}}}(w - i)}}{{iw{\psi _{\ln {S_T}}}( - i)}}} \right]} \;dw,
\end{equation}

\begin{equation}\label{pi2}
{\Pi _2} = \frac{1}{2} + \frac{1}{\pi }\int_0^\infty  {{\mathop{\rm Re}\nolimits} \left[ {\frac{{{e^{ - iw\ln (K)}}{\psi _{\ln {S_T}}}(w)}}{{iw}}} \right]} \;dw,
\end{equation}

\noindent where ${\psi _{\ln {S_T}}}$ is the characteristic function of the log-asset price and ${\mathop{\rm Re}\nolimits} [ \bullet ]$  denotes the real operator. European call prices can be obtained by first computing  $\Pi _1$ and $\Pi _2$, and then substituting these values into \eqref{CDPD}, whereas European puts can be determined through the put-call parity. We refer to \cite{Crisostomo2014} for a mathematical derivation and an implementation in MATLAB\textsuperscript{TM}.

In a comprehensive survey, \cite{Schmelzle2010a} concludes that the integrands in \eqref{pi1} and \eqref{pi2} decay rapidly and can be approximated through numerical integration. However, the DPD implementation faces three potential shortcomings:
\\
\begin{enumerate}
  \item[\textbf{1.}] \textbf{Discontinuities in the integrand functions}: The characteristic function of many stochastic volatility and jump-related processes contains a complex logarithm that may generate numerical instability. For instance, \cite{Schobel1999} give several examples where Heston's original characteristic function shows discontinuities and numerical integration may lead to incorrect option prices. This problem, however, can be circumvented in many models by an appropriate reformulation of the underlying characteristic function \cite{Albrecher2007, Lord2010}.
  \item[\textbf{2.}] \textbf{Singularity at $w=0$}: The DPD integrands are not defined at their lower integration limit. Lewis \cite{Lewis2001} analyzes this singularity and concludes that the integrands are finite as  $w$ tends to zero. Nevertheless, this divergence should be treated with caution, since inappropriate handling can result in pricing errors.
  \item[\textbf{3.}] \textbf{Number of evaluations}: To obtain option prices through the DPD, three characteristic function evaluations are required per integration point (two for $\Pi _1$ and another one for $\Pi _2$). Thus, if the integration grid is divided into $N$ points, $3N$ evaluations are needed per option priced or $3NM$ for a set of $M$  options. While this may not be a problem for occasional pricing, the CPU effort can become burdensome when calculating many option prices simultaneously or in real-time contexts.
\end{enumerate}

\subsection{Strike Vector Computations}\label{s:2.2}

Zhu \cite{Zhu2010} proposes a simple yet effective trick to reduce the computational effort of the DPD and other Fourier methods. The key insight is that the required characteristic function evaluations, both in $\Pi _1$   and $\Pi _2$, differ for each expiry, but are independent of the strike. Therefore, for a given  $T$, characteristic function values can be computed once and re-used to price options with different strikes. This idea can be implemented through vectorization or by a catching technique, as suggested by Kilin \cite{Kilin2011}.

Specifically, if we introduce a vector of strikes  ${\bf{K}}$ in the calculation of \eqref{pi1} and \eqref{pi2}, the probability vectors ${{\bf{\Pi }}_{\bf{1}}}$  and ${{\bf{\Pi }}_{\bf{2}}}$, are given by

\begin{equation}\label{VCDPD}
{{\bf{\Pi }}_{\bf{1}}} = \frac{1}{2} + \frac{1}{\pi }\int_0^\infty  {{\mathop{\rm Re}\nolimits} \left[ {\frac{{{e^{ - iw\ln ({\bf{K}})}}{\psi _{\ln {S_T}}}(w - i)}}{{iw{\psi _{\ln {S_T}}}( - i)}}} \right]} \;dw
\end{equation}

\begin{equation}\label{Vpi1}
{{\bf{\Pi }}_{\bf{2}}} = \frac{1}{2} + \frac{1}{\pi }\int_0^\infty  {{\mathop{\rm Re}\nolimits} \left[ {\frac{{{e^{ - iw\ln ({\bf{K}})}}{\psi _{\ln {S_T}}}(w)}}{{iw}}} \right]} \;dw.
\end{equation}

\noindent And, thus, the vector of call prices can be computed as
\begin{equation}\label{Vpi2}
{\bf{C}}(T,{\bf{K}}) = {S_0}{\rm{ }}{{\bf{\Pi }}_{\bf{1}}} - {e^{ - rT}}{\bf{K}}{\rm{ }}{{\bf{\Pi }}_{\bf{2}}}.
\end{equation}

\noindent Since the characteristic function evaluations are typically the most burdensome part of the calculations, vectorization significantly reduces the CPU effort while preserving two distinct advantages of the DPD: (i) the flexibility to choose any strikes and integration method and (ii) the intuitive probabilistic pricing \'a la Black-Scholes.

\subsection{Combining $\Pi _1$ and  $\Pi _2$ in a single integral}\label{s:2.3}

Attari \cite{Attari2004} proposes a DPD reformulation that calculates option prices through a single integral. Specifically, by exploiting the similarities in  $\Pi _1$  and $\Pi _2$, Attari's formula merges the integrands in \eqref{pi1} and \eqref{pi2} into a single pricing expression of the form

\begin{equation}\label{AT1}
C(T,K) = {S_0} - {\rm{ }}{e^{ - rT}}K\left( {\frac{1}{2} + \frac{1}{\pi }\;\int_0^\infty  {{I_A}(w)dw} } \right),
\end{equation}
\noindent where

\begin{equation}\label{AT2}
{I_A}(w) = \frac{{({\mathop{\rm Re}\nolimits} ({\psi _{\ln {S_T}}}(w)) + {\textstyle{{{\mathop{\rm Im}\nolimits} ({\psi _{\ln {S_T}}}(w))} \over w}})\cos (w\ln (K)) + ({\mathop{\rm Im}\nolimits} ({\psi _{\ln {S_T}}}(w)) - {\textstyle{{{\mathop{\rm Re}\nolimits} ({\psi _{\ln {S_T}}}(w))} \over w}})\sin (w\ln (K))}}{{1 + {w^2}}}
\end{equation}

\noindent Compared to the integrands in the DPD, ${I_A}(w)$ includes a quadratic term in the denominator, ensuring a faster decay rate. Furthermore, strike vectorizations can also be employed to speed up the computations, since Attari's characteristic function evaluations are independent of the strike.

\subsection{The COS method}\label{s:2.4}

Fang and Oosterlee \cite{Fang2009} introduce a pricing method based on Fourier-cosine expansions that offers a highly efficient way to recover the density of the underlying from the characteristic function. To benchmark our results to the BENCHOP project \cite{VonSydow2015}, we employ the COS method implementation developed by Ruitjter and Oosterlee \cite{Ruijter2015}, which derives the COS formula in three approximation steps: (i) express \eqref{C0} in terms of $\textrm{ln}({{S}_{t}}/K)$ and truncate the infinite integration range to an interval [a,b]; (ii) replace the density and option payoff by the first $N$ terms of their Fourier-cosine expansion and (iii) approximate the density-related coefficients using their characteristic function representation. Following these steps, the price of a European call can be obtained as:

\begin{equation}\label{COS1}
C(T,x)\approx {{e}^{-rT}}\sideset{}{'}\sum\limits_{n=0}^{N-1}\operatorname{Re}\left\{ \psi \left( \frac{n\pi }{b-a} \right){{e}^{in\pi \frac{x-a}{b-a}}} \right\}{{V}_{n}},
\end{equation}
\noindent with

\begin{equation}\label{COS2}
{{V}_{n}}=\frac{2}{b-a}\int_{a}^{b}{v(y,T)\ \cos \left( n\pi \tfrac{y-a}{b-a} \right)}dy,
\end{equation}

\noindent where $v(y,T)$ is the option payoff, $x=\textrm{ln}({{S}_{0}}/K)$, $y=\textrm{ln}({{S}_{T}}/K)$ and $\sum{}'$ indicates that the first summation term is weighted by one-half. In addition, to properly evaluate the performance of the COS method in multi-strike settings, we adapt the BENCHOP codes to simultaneously calculate option prices for different strikes.

\subsection{Carr-Madan's formula and fast Fourier transform}\label{s:2.5}

The FFT is an algorithm designed to compute Fourier transforms in an efficient way. Its application for option pricing was developed by Carr and Madan \cite{Carr1999}. The algorithm exploits periodicities and symmetries in the characteristic function evaluations to reduce the number of operations. For a given maturity, the FFT allows the simultaneous calculation of option prices for a variety of strikes.

\subsubsection{The modified call price}

Since the FFT can only be used in square-integrable functions, Carr-Madan's approach entails working with a modified call price where a dampening factor ${e^{\alpha \,\ln (K)}}$  is introduced to avoid the divergence at $w = 0$

\begin{equation}
{C_{\bmod }}(T,K) = {e^{\alpha\ln (K)}}C(T,K),
\end{equation}

\noindent where ${C_{\bmod }}(T,K)$  is the modified call price and $\alpha  > 0$  is the dampening parameter.  Using the Fourier inversion theorem, Carr-Madan's paper shows that the original call price can be recovered as:

\begin{equation}\label{CMformula}
C(T,K) = \frac{{{e^{ - \alpha \ln (K) - rt}}}}{\pi }\int_0^\infty  {{\mathop{\rm Re}\nolimits} \left[ {\frac{{{e^{ - iw\ln (K)}}{\psi _{\ln {S_T}}}(w - (\alpha  + 1)i)}}{{{\alpha ^2} + \alpha  - {w^2} + i(2\alpha  + 1)w}}} \right]} dw,
\end{equation}

\noindent where ${\psi _{\ln {S_T}}}$  is the characteristic function of the log-asset price.

\subsubsection{Integration with the fast Fourier transform}

Although \eqref{CMformula} can be directly used to compute call prices, it is common to evaluate it through the FFT. The FFT specifically computes sums of the form:
\begin{equation}
y(m) = \sum\limits_{n = 1}^N {{e^{ - i\frac{{2\pi }}{N}(m - 1)(n - 1)}}} x(n)\;\;\;\;for\,m = 1,....,N.
\end{equation}

\noindent Therefore, before applying the algorithm, the call price in \eqref{CMformula} should be expressed in the appropriate summation form. The first step is to approximate the integral by a grid of $N$  equidistant points, thus establishing an upper integration limit $N\Delta w$. Next, by setting the grid points as ${w_n} = (n - 1)\Delta w$, and using the trapezoidal rule, an \textit{individual} call price can be computed as

\begin{equation}
\mathord{\buildrel{\lower3pt\hbox{$\scriptscriptstyle\frown$}}
\over C} (K) \approx \sum\limits_{n = 1}^N {{e^{ - i{w_n}\,\ln (K)}}} f({w_n})\Delta w,
\end{equation}
\noindent where
\begin{equation}
f({w_n}) = \;{e^{\alpha \,\ln (K) - rT}}\frac{{{\psi _{\ln {S_T}}}({w_n} - (\alpha  + 1)i)}}{{{\alpha ^2} + \alpha  - {w_n}^2 + i(2\alpha  + 1)w}}.
\end{equation}

\noindent However, the FFT algorithm takes an $N$-sized vector $x(n)$ as input and returns another $N$-sized vector $y(m)$ as output. Consequently, the choice of $N$ determines both the number of strikes and the integration grid size. Furthermore, two constraints must be respected. First, the strikes must be placed at an equal distance in the log space\footnote{We define the strike grid as  ${k_m} =  - {k_{\max }} + (m - 1)\Delta k + \ln ({S_0})$ with $m = 1,...,N$ and $k = \ln K$. This choice entails setting the log FFT strikes symmetrically centered around $K = {S_0}$.}. Second, the Nyquist relation $\Delta k\Delta w = 2\pi /N$ must also be obeyed.

Putting all together, the prices of $N$ call options can be simultaneously obtained as

\begin{equation}
\mathord{\buildrel{\lower3pt\hbox{$\scriptscriptstyle\frown$}}
\over C} ({k_m}) \approx \sum\limits_{n = 1}^N {{e^{ - i\frac{{2\pi }}{N}(n - 1)(m - 1)}}} g({w_n})\,\,\,for\,m = 1,....,N,
\end{equation}
\noindent where
\begin{equation}\label{FFTsums}
g({w_n}) = \;{e^{ib{w_n} + \alpha \,{k_m} - rT}}\frac{{{\psi _{\ln {S_T}}}({w_n} - (\alpha  + 1)i)}}{{{\alpha ^2} + \alpha  - {w_n}^2 + i(2\alpha  + 1)w}}\Delta w.
\end{equation}

\noindent To harness the speed advantages of the FFT, the sums in \eqref{FFTsums} must be divided in two sequences: one with the odd terms and another with the even ones. The key insight is that the characteristic function evaluations required in the odd sequence are repeated for the even one. Thus, previously computed values can be used to reduce the number of operations. This strategy is reinforced by decomposing the odd and even sequences into two additional subsequences. And continuing this decimation until we obtain  $N/2$ subsequences of length 1, the FFT algorithm is able to reduce the computational effort from an order of ${N^2}$ to an order of $N\;{\rm{lo}}{{\rm{g}}_2}(N)$.

\subsubsection{FFT limits and alternatives}

The main FFT drawbacks stem from the restrictions imposed in the strike and integration grids:

\begin{enumerate}
  \item[\textbf{1.}] \textbf{Strike grid}. To achieve a fully efficient decimation the number of strikes must be a power of 2. Moreover, those ${2^d}$ strikes must be equidistantly placed in the log space. This means that the number and location of the resulting FFT prices will rarely match our needs. Prices closer to our strike needs can be computed by increasing $N$  or by interpolating across the prevailing strikes, but both strategies impact the merits of the FFT; a higher $N$  implies calculating more option prices than needed, whereas interpolation affects pricing accuracy.
  \item[\textbf{2.}] \textbf{Relationship between the strike and integration grid}. The constraint $\Delta k\Delta w = 2\pi /N$  imposes an inverse relationship between the integration step width and the output prices. Specifically, finer integration grids will lead to coarser strikes; thus, if we try to improve the pricing accuracy by reducing  $\Delta w$, the output prices will be more dispersed, increasing the need for interpolation.
  \item[\textbf{3.}] \textbf{Integration methods}. Since the FFT requires equidistant integration, only the most simple quadrature rules can be used to recover option prices. This compares unfavorably to other Fourier-based methods, where more efficient techniques can be employed to speed up the calculations.
\end{enumerate}

Carr and Madan \cite{Carr1999} suggest using an $N = 4096$. However, for most equity underlyings, there are rarely more than 20 or 30 actively traded strikes per maturity. Therefore, if we employ a much higher $N$, only a small fraction of the final FFT prices will fall within the usual trading ranges, which in turn means that many of these prices might be left unused\footnote{For example, out of the 4096 FFT prices calculated by Carr and Madan \cite{Carr1999}, only about 67 fall within the $\pm 20\%$ log-strike interval \cite{Chourdakis2005}.}.

To address these constraints, Chourdakis \cite{Chourdakis2005} introduces a Fractional FFT method (FRFT) that relax the restriction $\Delta k\Delta w = 2\pi /N$, providing more flexibility in the construction of the strike and integration grids. However, this method does not relax the requirement to place all the strike and integration points equidistantly, which is a fundamental FFT constraint.

Alternatively, the Carr-Madan formula in \eqref{CMformula} can be directly used to price call options without manipulation. Using a slightly modified version of \eqref{CMformula}, \cite{Matsuda2004} reports accurate option prices and negligible approximation errors for a variety of models. In addition, Carr-Madan's formula can be optimized through strike vectorization, since the characteristic function evaluations are independent of the strike.

\subsection{Numerical setup and error analyses}

We investigate the pricing biases and computational speed of seven pricing choices:

\begin{itemize}
\item \textbf{DPD}: Delta Probability Decomposition. Call values are individually computed through equations \eqref{CDPD} to \eqref{pi2}.
\item \textbf{DPD-OPT}: Optimized DPD. Strike vector computations are used to simultaneously compute call values for a variety of strikes. Equations \eqref{VCDPD} to \eqref{Vpi2} are used.
\item \textbf{AT-OPT}: Optimized Attari approach. Call values are computed with equations \eqref{AT1} and \eqref{AT2}. The CPU burden is optimized through strike vectorizations.
\item \textbf{COS-OPT}: Optimized COS method. A multi-strike version of \eqref{COS1} and \eqref{COS2} is used to calculate option prices. Following \cite{Fang2009}, the truncation range is obtained through the first four cumulants of $\textrm{ln}({{S}_{T}}/K)$ and a scale parameter $L$\footnote{We employ the truncation range $[a,b]=\left[ {{c}_{1}}-L\sqrt{{{c}_{2}}+\sqrt{{{c}_{4}}}},\ {{c}_{1}}+L\sqrt{{{c}_{2}}+\sqrt{{{c}_{4}}}} \right]$  where ${{c}_{n}}$  denotes the $n$-th cumulant of $\textrm{ln}({{S}_{T}}/K)$.}.
\item \textbf{FFT}: Standard FFT. Vector operations (instead of loops) are used to improve the performance. After experimenting with different values, we settle for an $\alpha = 1.75$, which delivers a ${10^{-10}}$  accuracy for all the models tested. Options that do not exactly fall in the FFT strike grid are exponentially interpolated.
\item \textbf{FFT-SA}: Strike-adjusted FFT. Call values are determined by successive FFT runs. Strike grids are adjusted to match all the required options in at least one FFT run, thus avoiding interpolation.
\item \textbf{CM-OPT}: Optimized Carr-Madan formula. Call values are computed using equation \eqref{CMformula} and strike vector computations.
\end{itemize}

These Fourier implementations can be subject to three forms of error:
\begin{enumerate}
  \item[\textbf{1.}] \textbf{Truncation error}: All methods require evaluating integrals in either $[0,\infty)$ or $(-\infty,\infty)$. To numerically approximate such integrals, the integration domain must be truncated by choosing appropriate integration limits, hence introducing a truncation error. For a given domain, the order of truncation errors can be different depending on (i) the underlying stochastic model and (ii) the Fourier implementation employed to obtain option prices. The rationale is that characteristic and density functions for different underlying models exhibit different decay rates, whereas the integrands described in subsections \ref{s:2.1} to \ref{s:2.5} also portray varying decay speeds \cite{Lee2004}.
  \item[\textbf{2.}] \textbf{Discretization error}: Except for the COS formula, the truncated integrals in all methods are evaluated through finite integration grids, thus introducing a sampling error. Different characteristic functions and Fourier implementations also affect the smoothness of the integrands, impacting discretization errors. To facilitate comparisons, in our analyses option prices are computed through the trapezoidal rule.

      In the COS method, the truncated integral is approximated by a finite number of Fourier-cosine expansion terms. Once a truncation domain has been chosen, discretization errors in the COS method depends on the decay rate of the cosine series coefficients. Therefore, the number of cosine terms used in the approximation can be compared to the integration size in quadrature-based approaches, since both determine the number of summation terms that are required to achieve a given accuracy.

  \item[\textbf{3.}] \textbf{Interpolation error}: This error arises when a pricing method does not provide the price for a desired strike. Consequently, in our setting, this error is specific to the FFT, since all the other variants can evaluate any required strike.
\end{enumerate}

To properly discriminate between these errors, all option prices in our \textit{accuracy comparisons} are calculated (i) with a high precision of ${10^{-10}}$, (ii) using common integration domains and (iii) using ${2^{d}}$ values for the integration grid and the number of terms in the Fourier-cosine expansion. Conversely, for the \textit{speed comparison}, the accuracy is set at a more practical ${10^{ - 4}}$ and we relax the common integration domain and ${2^{d}}$ constraint, thus allowing each method to optimize its pricing requirements. We compare how fast each method is able to price a variable number of options, covering a wide range of needs from 1 to 2500 options. Numerical calculations are performed using an Intel Core i7-3770 CPU @ 3.40GHz and 16 GB RAM.

\subsection{A first test with the BSM model}

We first apply all Fourier implementations to the BSM model, whose characteristic function is given by
\begin{equation}
\psi _{\ln ({S_t})}^{BSM}(w) = {e^{iw[{\rm{ln(}}{{\rm{S}}_0}) + (r - 0.5{\sigma ^2})t] - 0.5{w^2}{\sigma ^2}t}}.
\end{equation}

\subsubsection{Pricing accuracy in the BSM model}

We employ the parameters $S_0$ = 50, $\sigma = 0.25$ and r = 0.05. Accuracy is evaluated at six option configurations, spanning three different strikes $K = [30,  50, 70]$ and two maturities $T = [0.1, 1]$. The integration range is set at $w = (0, 100]$ for all Fourier methods except the COS formula, where an $L = 13$ is required to achieve an accuracy of ${10^{-10}}$. Reference values are computed through the BSM closed-form solution.

\begin{figure}
\begin{center}
\includegraphics[scale=0.5]{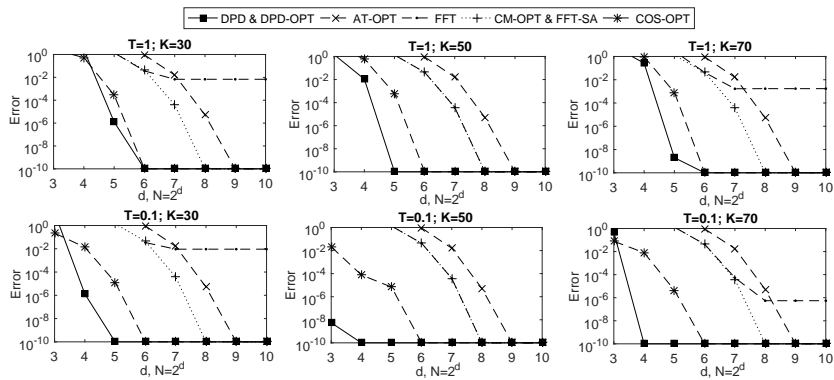}
\caption{Error convergence in the BSM model. Truncation range set to (0, 100] or $L = 13$ in the COS-OPT. Reference values: 21.5036288308, 6.1679994652 and 0.8986170065 for $T=1$, 20.1496256242, 1.7004462835 and 0.0000139309 for $T=0.1$.}
\label{Figure 1}
\end{center}
\end{figure}

As Figure~\ref{Figure 1} shows, most Fourier methods converge to the reference BSM values. The DPD and DPD-OPT achieve convergence with the smallest integration grids --between 16 and 64 points--, whereas the COS method requires an $N = 64$ in all option configurations. On the other hand, the AT-OPT suffers the highest discretization errors, requiring  $N = 512$ points to deliver an accuracy of ${10^{-10}}$.

The FFT converge to the reference values in the ATM options, but a single grid cannot exactly match all the required strikes; thus the OTM and ITM option prices have been exponentially interpolated, introducing an interpolation error. In contrast, both the CM-OPT and FFT-SA deliver a ${10^{-10}}$ accuracy for all strikes and maturities with $N= 256$ points\footnote{The CM-OPT and FFT-SA rely on the same pricing approach and can evaluate any specific strike. Therefore, when the same truncation range and integration grid is used, they are equivalent in terms of accuracy}.

Summing up, both truncation and discretization errors are small and easy to manage in the BSM model, and high precision values can be obtained integrating in $w = (0, 100]$ or with $L= 13$. These results derive from the well-behaved diffusive properties of the geometric Brownian motion, which in turn entails a smooth and rapidly decaying characteristic function.

\subsubsection{Computational speed in the BSM model}

To investigate the CPU effort, we obtain the truncation ranges required to attain full convergence and the number of points that deliver an accuracy of ${10^{-4}}$. Reported times are calculated by averaging the computational effort in 100 independent runs.

As Table~\ref{Table 1} shows, the multi-strike version of the COS method is faster than any other alternative, being on average 4, 9 and 15 times faster than the DPD-OPT, CM-OPT and AT-OPT respectively. Note that the COS-OPT is faster that the DPD-OPT despite requiring a higher $N$, thus highlighting the more efficient computation of the cosine series terms compared to the grid point evaluations in quadrature-based methods.

In contrast, the two slowest methods are the unoptimized DPD --which perform separate computations for each option-- and the FFT. Leaving interpolation biases aside, the FFT requires 128 points to deliver a ${10^{-4}}$ accuracy. Therefore, this method always computes a minimum of 128 option prices, impacting its performance when fewer prices are required. The FFT efficiency improves with the number of options, but its speed is still notably lower than in any strike-optimized alternative.

Against the FFT constraints, the CM-OPT offers three advantages. It allows: (i) pricing any number of strikes (ii) avoiding interpolation biases and (iii) achieving a ${10^{-4}}$  accuracy with a lower $N$. As a result, the CM-OPT is simultaneously faster and more accurate than the FFT, rendering the latter inefficient. Following these figures, we decided not to pursue the speed comparison for the FFT-SA, which requires at least twice the FFT's computing times and cannot improve the CM-OPT accuracy.

\begin{table}
\tbl{CPU times required to achieve a ${10^{-4}}$  accuracy in the BSM model [milliseconds].}
{\begin{tabular}[l]{@{}lcccccccc}\toprule
        &           &           & & & \multicolumn{4}{l}{N. of options priced}                      \\
\colrule
Method  & Domain   & Minimum N & 1       & 10      & 25      & 100      & 500      & 2500      \\
\colrule
DPD     & $(0, 89{]}$ & 26        & 0.176 & 1.791 & 4.373 & 17.36 & 88.44 & 435.1 \\
DPD-OPT & $(0, 89{]}$ & 26        & 0.176 & 0.232 & 0.279 & 0.357  & 1.088  & 3.634   \\
AT-OPT  & $(0, 79{]}$ & 173       & 0.156 & 0.244 & 0.328 & 0.874  & 2.690  & 18.10  \\
FFT     & $(0, 77{]}$ & 128       & 0.524 & 0.524 & 0.524 & 0.524  & 5.781  & 317.9 \\
CM-OPT  & $(0, 77{]}$ & 97        & 0.104 & 0.154 & 0.194 & 0.504  & 1.682  & 11.29  \\
COS-OPT  & $L = 13$ & 37        & 0.011 & 0.022 & 0.030 & 0.062  & 0.235  & 1.109  \\
\botrule
\end{tabular}}
\label{Table 1}
\end{table}

\section{The Bates Jump-diffusion Model}\label{s:3}

\subsection{Model description}

Bates \cite{Bates1996a} proposes a modelling framework which blends the Heston model with lognormally distributed price jumps. Under the risk-neutral measure, the Bates dynamics are given by

\begin{equation}
\begin{aligned}
  d{S_t} = (r - \lambda {\mu _J}){S_t}dt + \sqrt {{V_t}} {S_t}dW_t^1 + {J_t}{S_t}d{N_t} \\
  d{V_t} = a(\bar V - {V_t})dt + \eta \sqrt {{V_t}} dW_t^2,
\end{aligned}
\end{equation}

\noindent where ${S_t}$ is the price of the underlying asset at time $t$ , $r$  the risk free rate, ${V_t}$ the variance at time $t$, $\bar V$ the long-term variance, $a$  the variance mean-reversion speed,  $\eta$ the volatility of the variance process and $W_t^1$, $W_t^2$  are two Weiner processes with correlation $\rho$. In addition, ${N_t}$  is a Poisson process with intensity $\lambda$, and ${J_t}$  are the jump sizes, which are lognormally distributed with an average jump size ${\mu _J}$  and standard deviation ${v_J}$. Therefore, conditional on a jump occurring, the logarithm of the jump size is normally distributed with parameters

\begin{equation}
\ln (1 + {J_t}) \sim N\left( {\ln (1 + {\mu _J}) - \frac{{v_J^2}}{2},{v_J}} \right).
\end{equation}

The rationale for mixing stochastic volatility and jumps is based on empirical grounds. Evidence shows that volatilities can change drastically over time and that asset prices experience price jumps. As a result, both observed returns and market expectations are characterized by distributions that exhibit substantial asymmetries and fat-tails, particularly in the short-term \cite{Cont2001}.

Most empirical studies support the main features of the Heston model --mean-reverting volatility and correlated volatility and asset shocks--, concluding that Heston dynamics provide a good fit to the prices of long-term options \cite{Bakshi1997a, Crisostomo2014}. However, the diffusive behavior of the Heston model struggles to generate the leptokurtic distributions that are routinely implied by short-dated options \cite{Jones2003, Sepp2003}. Conversely, as explained in \cite{Carr2003}, lognormal jumps can significantly contribute to explaining the price of short-term options, but their smile effects flatten out quickly in longer time periods.

Consequently, by combining stochastic volatility and lognormal jumps, the Bates model offers a versatile modelling scheme that can be used to accommodate both the short and the long end of the volatility surface.

\subsection{Bates characteristic function}

Since the lognormal jumps are statistically independent from the stochastic volatility dynamics, the Bates characteristic function can be obtained by multiplying its individual components
\begin{equation}
\psi _{\ln ({S_t})}^{Bates}(w) = \;\psi _{\ln ({S_t})}^{Heston}(w)\;.\;\psi _{\ln ({S_t})}^{Jump}(w).
\end{equation}

For the Heston model, we follow the formulation in \cite{Gatheral2006}, which is free of the complex logarithm problem mentioned in section 2 \cite{Lord2010}. For the lognormal jump, we use the derivation in \cite{Schoutens2004b}. Multiplying and rearranging terms yields

\begin{equation}
\psi _{\ln ({S_t})}^{Bates}(w) = {e^{[C(t,w)\bar V + D(t,w){V_0} + J(t,w) + iw\ln ({S_0}{e^{(r - \lambda {\mu _J})t}}]}},
\end{equation}

\noindent with

\begin{equation}
\begin{aligned}
C(t,w) = \;a\left[ {{r_ - } \cdot t - \frac{2}{{{\eta ^2}}}ln\left( {\frac{{1 - g{e^{ - ht}}}}{{1 - g}}} \right)} \right]; \\
D(t,w) = {r_ - }\frac{{1 - {e^{ - ht}}}}{{1 - g{e^{ - ht}}}}\\
J(t,w) = \lambda t\left[ {{{\left( {1 + {\mu _J}} \right)}^{iw}}{e^{{\textstyle{1 \over 2}}v_J^2iw(iw - 1)}} - 1} \right];\\
{r_ \pm } = \frac{{\beta  \pm h}}{{{\eta ^2}}}\;;\;\;h = \sqrt {{\beta ^2} - 4\alpha \gamma } ;\,\,\,\,g = \frac{{{r_ - }}}{{{r_ + }}}\;\;\;\\
\alpha  =  - \frac{{{w^2}}}{2} - \frac{{iw}}{2}\;;\;\;\beta  = a - \rho \eta iw\;;\;\;\gamma  = \frac{{{\eta ^2}}}{2},
\end{aligned}
\end{equation}

\noindent where $C(t,w)\bar V$ and $D(t,w){V_0}$  come from the Heston model, $J(t,w)$  is a jump-specific component, while $iw\ln ({S_0}{e^{(r - \lambda {\mu _J})t}})$  accounts for the combined risk-neutral drift.

\subsection{Numerical results}

\subsubsection{Pricing accuracy in the Bates model}

The parameter set is taken from \cite{Duffie2000b}: ${S_0} = 100$, ${V_0} = 0.008836$, $\bar V  = 0.014$, $a = 3.99$, $\eta = 0.27$, $r = 0.0319$, $\rho= -0.79$, $\lambda = 0.11$,  ${\mu _J}= -0.12 $ and ${v_J} = 0.15$. The accuracy is evaluated at three strikes $K = [60, 100, 140]$ and two tenors $T = [0.1, 1]$. Due to the jump component, the Bates characteristic function exhibits fatter tails than in the BSM model, thus increasing truncation error. We find that in order to achieve a ${10^{-10}}$  accuracy, the integration range needs to be expanded to $w = (0, 500]$ in most quadrature-based methods, whereas an $L = 30$ is required in the COS formula\footnote{Following \cite{Fang2009}, we compute the truncation range in the Bates model through the first two cumulants of $\textrm{ln}({{S}_{T}}/K)$. This choice, however, leaves the $4$th-cumulant $c_4$ out of the calculation, thus resulting in larger values of $L$.}. Working in these domains, reference values are obtained through the concurrent prices of the AT-OPT and the CM-OPT, integrating with ${10^{6}}$ points.

\begin{figure}
\begin{center}
\includegraphics[scale=0.5]{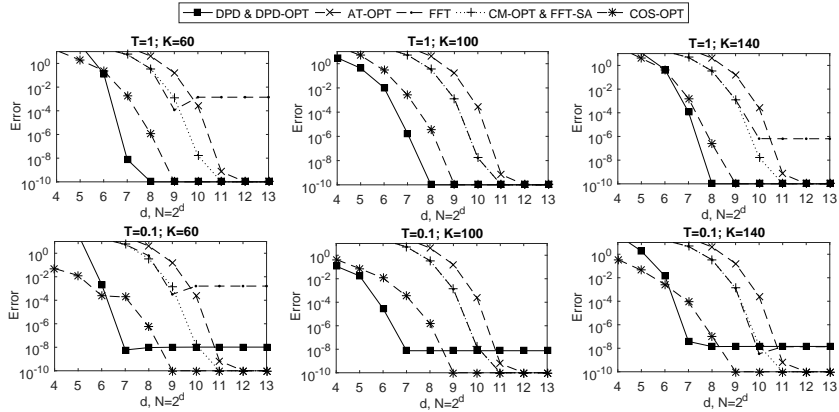}
\caption{Error convergence in the Bates model. Truncation range set to (0, 500] or $L = 30$ in the COS-OPT. Reference values: 441.9030506459, 6.7577754525 and 0.0058803882 for $T=1$, 40.1913714101, 1.4817911043 and 0.0000688740 for $T=0.1$.}
\label{Figure 2}
\end{center}
\end{figure}

As Figure~\ref{Figure 2} shows, the DPD and DPD-OPT offer the fastest convergence rate for the $T=1$ options. In contrast, the pricing biases observed in the $T=0.1$ can be attributed to truncation errors, and an expanded domain (0, 649] is required to eliminate the remaining $O({10^{-8}})$ errors. These results highlight: (i) the larger truncation error of the DPD integrands compared to other methods and (ii) the slower decay of the Bates characteristic function in short expiries.

The COS-OPT ranks as second-best, delivering a 10 digits accuracy with $N = 2^9$. Note, however, that for the accuracy levels of ${10^{-2}}$ or below, the COS formula converge to the ITM and OTM prices with the smaller $N$, thus becoming the most efficient method. In contrast, the AT-OPT and the Carr-Madan variants exhibit the slowest convergence, requiring between $2^{11}$ and $2^{12}$ points to achieve a ${10^{-10}}$ accuracy\footnote{Except for the ITM and OTM strikes in the FFT, where interpolation errors result in biases of $O({10^{-3}})$ or lower}.

Overall, except for small interpolation or truncation biases, no major problems are observed in the Bates model, and high precision values can be obtained integrating in $w$ = (0, 500] or with $L = 30$. However, these results expose the increased complexity of the Bates model compared to the BSM framework, which entails (i) fatter tails due to a slower decaying characteristic function, increasing truncation errors and (ii) a less smooth probabilistic distribution, increasing discretization errors.

\subsubsection{Computational speed in the Bates model}

To evaluate the CPU burden, we obtain the truncation ranges required for full convergence and the number of points that deliver a ${10^{-4}}$ accuracy. As expected, the larger domains and sampling frequencies impact the required $N$. Computational times in the Bates model are, on average, 6 times higher than in the BSM framework.

Table~\ref{Table 2} shows that the COS-OPT is again the fastest alternative for all pricing needs. The multi-strike version of the COS method is roughly 3 times faster than the DPD-OPT and 7 times higher than the CM-OPT. The COS-OPT efficiency stems from (i) the smaller truncation range and fast decay of the cosine series coefficients --both resulting in a lower $N$-- and (ii) the less burdensome computations of the Fourier series terms compared to quadrature evaluations in other methods.

Among quadrature-based method, the CM-OPT and FFT stands out for minimizing truncation errors, only requiring a Fourier domain of size (0 470]. However, the FFT requires $N = 1024$ point to achieve a ${10^{-4}}$ accuracy, thus always computing a minimum of 1024 option prices. In contrast, the CM-OPT can price any strikes, avoid interpolation biases and achieves the same accuracy with a smaller $N$, thus delivering faster and more accurate option prices than the FFT. Finally, despite requiring the largest $w$-range, the DPD and DPD-OPT only need 176 points to achieve a ${10^{-4}}$ accuracy (0.27 points per unit of $w$), thus offering a remarkable sampling efficiency.

\begin{table}
\tbl{CPU times required to achieve a ${10^{-4}}$  accuracy in the Bates model [milliseconds].}
{\begin{tabular}[l]{@{}lcccccccc}\toprule
        &           &           & & & \multicolumn{4}{l}{N. of options priced}                      \\
\colrule
Method  & Domain   & Minimum N & 1       & 10      & 25      & 100      & 500      & 2500      \\
\colrule
DPD     & (0, 649{]} & 176       & 0.491  & 4.928  & 12.35 & 49.53  & 246.5 & 1232 \\
DPD-OPT & (0, 649{]} & 176       & 0.491  & 0.698  & 0.827  & 1.812  & 5.214   & 35.49  \\
AT-OPT  & (0, 478{]} & 1091      & 1.398  & 2.327  & 2.793  & 4.999  & 24.98  & 129.3 \\
FFT     & (0, 470{]} & 1024      & 24.60 & 24.60 & 24.60 & 24.60 & 24.60  & 303.6 \\
CM-OPT  & (0, 470{]} & 622       & 0.330  & 0.553  & 1.073  & 2.416  & 14.33  & 77.89  \\
COS-OPT  & $L = 30$ & 164        & 0.155 & 0.216 & 0.254 & 0.621  & 1.846  & 11.21  \\
\botrule
\end{tabular}}
\label{Table 2}
\end{table}

\section{The Asymmetric Variance Gamma}\label{s:4}

\subsection{Model description}

The Variance Gamma model was introduced in \cite{Madan1990a}. However, it is its asymmetric version \cite{Madan1998a} which has achieved the greatest acceptance. The AVG is a purely discontinuous process where the underlying asset evolves through a combination of many small jumps and a limited number of big jumps. The risk-neutral AVG dynamics are given by:

\begin{equation}
{S_t} = {S_0}{e^{(r + \lambda )t + X(t;\sigma ,v,\theta )}},
\end{equation}

\noindent with
\begin{equation}
\lambda  = \frac{1}{v}\ln (1 - \theta v - \frac{{{\sigma ^2}v}}{2}),
\end{equation}

\noindent and
\begin{equation}
X(t;\sigma ,v,\theta ) = \theta G(t;v) + \sigma G(t;v){W_t},
\end{equation}

\noindent where $G(t;v)$  is a Gamma distribution with mean $t$  and variance $vt$, and ${W_t}$ is a Weiner process $N(0,1)$. Besides the risk free rate $r$, the model has three free parameters: $\sigma  > 0$ , $v > 0$  and $\theta \in \mathbb{R}$. In broad terms, $\sigma$ governs the volatility, $\theta$ the skewness, and $v$ provides control over the kurtosis. However, barring exceptions\footnote{For instance, when $\theta = 0$ the AVG distribution is symmetric, and $v$  alone determines the excess kurtosis, which is equal to  $3(1 + v)$.}, it is the particular combination of these three parameters which jointly determines the higher moments of the AVG distribution. We refer to \cite{Fiorani2004} for a detailed statistical characterization.

Among jump models, the AVG model offers one the most parsimonious approaches that can consistently price options with different moneyness and maturities. Furthermore, several empirical studies have shown that the AVG dynamics provide a very good fit to the observed equity returns \cite{Rebonato2004, Goncu2016}.

\subsection{AVG characteristic function}

The AVG model exhibits a closed-form solution for the valuation of European options. However, its numerical implementation entails working with Bessel functions of the second type, making it complex and numerically unstable \cite{Matsuda2004}.

Alternatively, the characteristic function of the AVG model is given by

\begin{equation}
\psi _{\ln ({S_t})}^{AVG}(w) = {\left( {\frac{1}{{1 - i\theta vw - ({\sigma ^2}v/2){w^2}}}} \right)^{t/v}},
\end{equation}
\noindent and can be directly used to calculate option prices through the Fourier methods presented in section \ref{s:2}. Some popular choices, however, can blow up for certain AVG parameter values \cite{Itkin2010}. We investigate these failures in the next section.

\subsection{Numerical results}\label{s:4.3}

\subsubsection{Pricing accuracy in the AVG model}

\vspace{3mm}
\noindent \textbf{Parameters based upon Madan, Carr, and Chang (1998)}
\vspace{3mm}

\begin{figure}
\begin{center}
\includegraphics[scale=0.5]{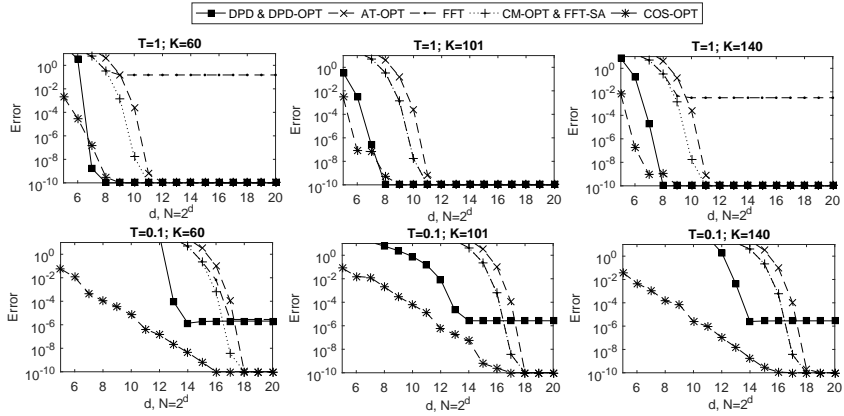}
\caption{Error convergence in the AVG model. Parameters based upon Madan, Carr, and Chang (1998). Truncation range set to $(0, 500]$ or $L = 8$ for $T = 1$, and $(0, 60000]$ or $L = 12$ for $T =0.1$. Reference values: 45.7164396686, 10.9815614276 and 0.1019706457 for $T=1$, 40.5972193355, 1.3938439616 and 0.0000061410 for $T=0.1$.}
\label{Figure 3}
\end{center}
\end{figure}

For our \textit{first pricing} test we employ the parameters ${S_0} = 100$, $\sigma = 0.12136$, $\theta  = -0.1436$,$v  = 0.3$ and $r  = 0.1$. We consider three strikes $K = [60, 101, 140]$ and two tenors $T = [0.1, 1]$. Despite its simple mathematical form, the slow hyperbolic decay of the AVG characteristic function complicates its numerical implementation. Specifically, to obtain a ${10^{-10}}$ accuracy across most pricing methods, the required integration range stands at  $w = (0, 500]$ or $L =8$ for the options at $T = 1$, but significantly increases to  $w = (0, 60000]$ or $L = 12$ for those at $T =0.1$. Working in these domains, reference values can be obtained through the concurrent prices of the CM-OPT and AT-OPT, integrating with ${10^{6}}$ points\footnote{This methodology replicates the numerical results in \cite{Ribeiro2004} and the related example in \cite{Hirsa2012}, for which \cite{Lewis2013} reports a high precision value of 11.3700278104.}. Figure~\ref{Figure 3} shows the convergency for all options.

For the  $T = 1$ options, the convergence pattern is relatively similar to that of the Bates model: the DPD-OPT achieves a ${10^{-10}}$ accuracy with the smallest $N$, whereas the COS-OPT is the most efficient for accuracies of ${10^{-7}}$ or lower.

In contrast, all Fourier methods require a much higher $N$ in the $T = 0.1$ options. When truncation errors play a prominent role, the COS formula converges significantly faster than any other alternative. This is particularly striking for the accuracy levels of ${10^{-1}}$ to ${10^{-6}}$, where the COS-OPT converges with a maximum $N = 2^{11}$ whereas other methods require much larger grids. The difference in $N$ can be attributed to both the smaller COS-OPT integration domain and the rapid decay of the cosine series coefficients.

Note that despite using a wide (0, 60000] range, the DPD and DPD-OPT fail to provide full convergence in the $T=0.1$ options. To eliminate the remaining biases, the integration limit must be expanded up to $w = 2.2 \cdot {10^7}$, evidencing a remarkably slow decay. The slowest convergence rate is observed in the AT-OPT and the Carr-Madam variants, both requiring a grid of size $N = 2^{18}$ to attain full accuracy.

\vspace{15mm}
\noindent \textbf{Parameters based upon Itkin (2010): Two problematic cases}
\vspace{3mm}

For our \textit{second pricing} test we consider a parametrization ${S_0} = 100$, $\sigma = 1$, $\theta  = 2$, $v  = 0.5$ and $r = 0.02$. We evaluate again three strikes  $K = [60, 90, 140]$ and two tenors $T = [0.1, 1]$. Convergency problems surface immediately when calculating the AVG reference values. Despite substantially increasing the truncation domain and sampling frequencies, we were unable to obtain concurrent option prices for any two Fourier methods. Furthermore, Table~\ref{Table 3} shows that most pricing choices blow up under this parameter set, producing negative or unfeasible call values.

The problem, according to \cite{Itkin2010}, can be traced down to the inequality constraint

\begin{equation}\label{VGRN}
 {{v}^{-1}}>\theta +0.5{{\sigma }^{2}}
\end{equation}

\noindent which must be respected in order to obtain a valid risk-neutral measure. However, it is remarkable that, despite being in a region where \eqref{VGRN} is not obeyed, the AT-OPT still delivers feasible option prices. In contrast to other methods, the AT-OPT produces call prices that are: (i) within reasonable positive bounds, (ii) monotonically increasing with time and (iii) monotonically decreasing across strikes.

\vspace{3mm}
\begin{table}[H]
\tbl{AVG pricing results for a parameter set where \eqref{VGRN} is not respected$^{\rm a}$.}
{\begin{tabular}[l]{@{}lcccccccc}\toprule
                             &      &  & $T = 0.1$ & & & $T=1$ &                      \\
\colrule
Method                       & $N$    & $K= 60$         & $K=90$          & $K= 140$         & $K =60$         & $K=90$          & $K=140$          \\
\colrule
DPD-OPT   & ${2^{24}}$& $1.08\mathrm{e}{+20}$  & $1.08\mathrm{e}{+20}$  & $1.08\mathrm{e}{+20}$ & -10.563 & -23.693 & -44.258 \\
AT-OPT          &${2^{24}}$ & 51.053 & 34.141 & 14.595 & 68.604  & 54.609  & 32.285  \\
CM-OPT \& FFT    & ${2^{24}}$  & -0.7811 & 0.4285  & 0.3253  & -$0.44\mathrm{e}{\text{-}4}$    & -$0.22\mathrm{e}{\text{-}4}$    & -$0.10\mathrm{e}{\text{-}4}$  \\
COS-OPT        & ${2^{24}}$ & $2.09\mathrm{e}{+6}$    & $3.01\mathrm{e}{+6}$    & $4.48\mathrm{e}{+6}$    & $1.03\mathrm{e}{+15}$ & $1.52\mathrm{e}{+15}$    & $2.32\mathrm{e}{+15}$ \\
\botrule
\end{tabular}}
\tabnote{$^{\rm a}$  Truncation range set to $(0, 1200000]$ or $L =12$ for the COS-OPT.}
\label{Table 3}
\end{table}
\vspace{3mm}

Finally, we explore a \textit{third parametrization} where ${S_0}= 100$, $\sigma = 1$, $\theta = 1.5$, $v  = 0.2$ and $r = 0.02$, thus surveying a region where inequality \eqref{VGRN} is respected.  Reference values for this parameter set can be obtained through the concurrent prices of the AT-OPT and DPD-OP, integrating in $w$ = (0, 1200000) with $N = 10^{8}$ points.

An outstanding result in this region is the failure of all the Carr-Madan variants with our standard numerical setup. Such failure arises due to a singularity that appears after substituting the AVG characteristic function into Carr-Madan's integrand \cite{Itkin2010}. Drilling down, we find that the blow ups are connected to the specific values of the dampening parameter $\alpha$. As reported in \cite{Carr1999}, in order to keep the AVG characteristic function finite, the choice of $\alpha$ should respect

\begin{equation}
\alpha  < \sqrt {\,\frac{{{\theta ^2}}}{{{\sigma ^4}}} + \frac{2}{{{\sigma ^2}v}}}  - \,\;\frac{\theta }{{{\sigma ^2}}} - \;1.
\end{equation}

\noindent thus requiring an $\alpha <1$ in our third AVG parametrization.

\begin{figure}[H]
\begin{center}
\includegraphics[scale=0.5]{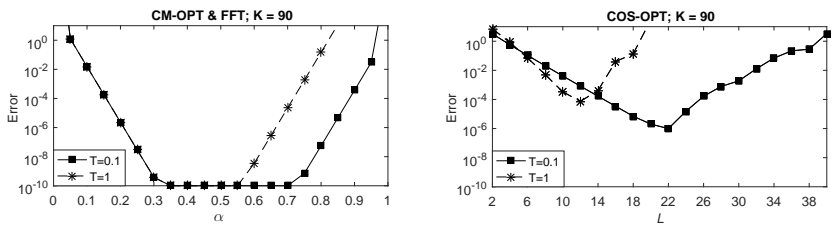}
\caption{Error convergence in the AVG model for a parameter set where \eqref{VGRN} is respected. Prices obtained with $N = {2^{24}}$. Truncation range set to $(0, 1200000]$ for the CM-OPT $\&$ FFT. Reference values: 58.9490408593 for $T=1$ and 20.0293202541 for $T=0.1$.}
\label{Figure 4}
\end{center}
\end{figure}

As a result, our initial choice $\alpha = 1.75$ fails to provide reasonable option prices. However, simply employing an $\alpha$ within the $(0, 1)$ range may also generate substantial mispricings. As Figure~\ref{Figure 4} shows, to achieve full accuracy across all option configurations, $\alpha$  must be specifically chosen in the range [0.35, 0.55], thus notably restricting its optimal values. The pricing biases increase for any $\alpha$ outside this range and, even within the feasibility region, both the FFT and the CM-OPT blow-up as $\alpha$ approaches  0 or 1.

\vspace{3mm}

Similarly, the COS formula also requires careful treatment. For this parameter set, Figure~\ref{Figure 4} shows that the COS-OPT is unable to achieve a ${10^{-10}}$  accuracy regardless of the choice of $L$. The maximum accuracy stands at ${10^{-6}}$ for $T=1$ and $L=22$, whereas only a lower ${10^{-4}}$ convergence is attained for $T=0.1$  with $L = 11$. In contrast, neither the DPD nor the AT-OPT suffer these problems. The error pattern is the same for different strikes.

\subsubsection{Computational speed in the AVG model}

For the speed tests we consider again the parametrization in \cite{Madan1998a}, where all methods are blow-up free and can be compared on an equal basis. Since truncation errors are remarkably different for the distinct expiries, we report separate comparisons for $T = 1$ and $T = 0.1$. Tables~\ref{Table 4} and \ref{Table 5} show the results.

\begin{table}
\tbl{CPU times required to achieve a ${10^{-4}}$  accuracy in the AVG model for $T = 1$ [milliseconds].}
{\begin{tabular}[l]{@{}lcccccccc}\toprule
        &           &           & & & \multicolumn{4}{l}{N. of options priced}                      \\
\colrule
Method  & Domain   & Minimum N & 1       & 10      & 25      & 100      & 500      & 2500      \\
\colrule
DPD     & (0, 462{]} & 114       & 0.253 & 2.539 & 6.327 & 25.33 & 126.3 & 632.1 \\
DPD-OPT & (0, 462{]} & 114       & 0.253 & 0.374 & 0.490 & 1.071  & 3.342   & 21.90  \\
AT-OPT  & (0, 334{]} & 773       & 0.475 & 0.801 & 1.300 & 2.900  & 15.84  & 90.31  \\
FFT     & (0, 295{]} & 512       & 5.868 & 5.868 & 5.868 & 5.868  & 5.868   & 320.8  \\
CM-OPT  & (0, 295{]} & 388       & 0.224 & 0.344 & 0.866 & 1.717  & 8.695   & 47.85   \\
COS-OPT  & $L = 8$ & 60       & 0.038 & 0.063 & 0.078 & 0.152  & 0.786  & 3.717  \\
\botrule
\end{tabular}}
\label{Table 4}
\end{table}

For the $T = 1$ options, the results are similar to those of the Bates model: the COS-OPT is the fastest method overall, whereas the DPD-OPT ranks as second-best. In contrast, the FFT and the unoptimized DPD are the slowest alternatives, with the FFT being on average 12 and 76 times slower than the DPD-OPT and the COS-OPT.

Conversely, for the $T = 0.1$ options, the slow AVG hyperbolic decay notably impacts computational times. To cope with the truncation error increase, we compute the integration ranges that deliver a ${10^{-6}}$ accuracy (instead of the usual ${10^{-10}}$), and then obtain the number of points that achieve an accuracy of ${10^{-4}}$.

\begin{table}
\tbl{CPU times required to achieve a ${10^{-4}}$  accuracy in the AVG model for $T = 0.1$ [milliseconds].}
{\begin{tabular}[l]{@{}lcccccccc}\toprule
        &           &           & & & \multicolumn{4}{l}{N. of options priced}                      \\
\colrule
Method  & Domain   & Minimum N & 1       & 10      & 25      & 100      & 500      & 2500      \\
\colrule
DPD     & (0, 79980{]} & 14889     & 4.257   & 45.75  & 115.1 & 461.0 & 2305 & 11534 \\
DPD-OPT & (0, 79980{]} & 14889     & 4.257   & 13.20   & 30.55  & 126.3 & 631.7 & 3130 \\
AT-OPT  & (0, 6997{]}  & 16041     & 5.567   & 11.13  & 19.95  & 71.65  & 347.8 & 1786 \\
FFT     & (0, 6536{]}  & 16384     & 5223 & 5223 & 5223 & 5223 & 5223 & 5223 \\
CM-OPT  & (0, 6536{]}  & 8625      & 1.571   & 3.665   & 9.429   & 39.63  & 205.7 & 1024  \\
COS-OPT  & $L = 12$ & 870      & 0.191 & 0.318 & 0.846 & 2.066  & 12.64  & 58.53  \\
\botrule
\end{tabular}}
\label{Table 5}
\end{table}

Due to the significantly larger truncation domains and $N$ values, the waiting times for the $T = 0.1$ options are, on average, 51 times higher than in the $T = 1$ expiries and 34 times higher than in the Bates model. Furthermore, since the truncation range increase is most prominent in the Fourier space, the COS method --which operates, in contrast, on the scaled density space-- delivers even better results than in previous comparisons, being 17 times faster than the CM-OPT. In contrast, the sluggish DPD/DPD-OPT decay and the low sampling efficiency of the AT-OPT result in notably higher integration sizes, making these methods particularly inefficient for the $T=0.1$ options.

\section{Conclusions}\label{s:5}

This paper analyses the speed and accuracy of seven Fourier-based pricing choices. We show that truncation errors increase as we move from the BSM to the Bates model and further intensify under the AVG dynamics. Discretization errors also increase when discontinuous jumps are considered, but the rise is modest and remains similar for both jump models.

Our analyses demonstrate the higher efficiency of strike vector computations compared to other traditional choices. In our tests, computing option prices through the AT-OPT, CM-OPT, DPD-OPT and COS-OPT is, on average, 54, 67, 165 and over 1500 times faster than in the FFT. We show that the multi-strike version of the COS formula is the overall fastest alternative, a result that stems from the lower truncation range required in COS method and the rapid decay of the cosine series coefficients.

We find that among quadrature-based methods: (i) the DPD-OPT exhibits the highest sampling efficiency but also the slowest decay rate, (ii) the CM-OPT stands out for minimizing truncation errors in the Fourier space and (iii) the AT-OPT suffers the largest discretization errors, requiring higher values of $N$ to achieve the same level of accuracy. As a result, the DPD-OPT performs best when pricing a high number of options, the CM-OPT is more efficient when only a few  prices are required, while the AT-OPT typically ranks as the slowest strike-optimized alternative.

We show that obtaining accurate option values can be particularly challenging in the AVG model. While all methods convergence under the BSM and Bates dynamics, large truncation errors significantly complicate the practical AVG implementation. Moreover, depending on the AVG parameters, specific Fourier implementations may completely fail to provide reasonable option prices: both the FFT and the CM-OPT can blow up in regions where inequality \eqref{VGRN} is respected, whereas the DPD-OPT and COS-OPT also fail when \eqref{VGRN} is not obeyed. In contrast, the AT-OPT seems to work fine for any AVG parameter values.

Finally, the comparison between the FFT and the CM-OPT deserves a special mention. While both are based on the same pricing approach, the CM-OPT's flexibility allows (i) pricing any required strikes, (ii) choosing any integration grid and (iii) avoiding interpolation biases. As a result, the CM-OPT is both faster and more accurate than the FFT, thus rendering this method inefficient. Based on our results, we see no reason to employ the FFT over the CM-OPT, but further analysis may be needed in order to confirm this hypothesis.

\section*{Acknowledgements}

We thank the anonymous referees and the Editor for their quick review and useful comments.

\bibliographystyle{gCOM}
\bibliography{libraryRCA}

\begin{thebibliography}{10}
\newcommand{\noopsort}[1]{}
\newcommand{\printfirst}[2]{#1}
\newcommand{\singleletter}[1]{#1}
\newcommand{\switchargs}[2]{#2#1}
\providecommand{\url}[1]{\normalfont{#1}}
\providecommand{\urlprefix}{Available at }

\bibitem{Albrecher2007}
H. Albrecher, P. Mayer, W. Schoutens, and J. Tistaert, \emph{{The little Heston
  trap}}, Wilmott Mag.  (2007), pp. 83--92.

\bibitem{Attari2004}
M. Attari, \emph{{Option Pricing Using Fourier Transforms: A Numerically
  Efficient Simplification}}, SSRN Electron. J.  (2004), pp. 1--7.

\bibitem{Bakshi1997a}
G. Bakshi, C. Cao, and Z. Chen, \emph{{Empirical Performance of Alternative
  Option Pricing Models}}, J. Finance 52 (1997), pp. 2003--2049.

\bibitem{Bakshi2000}
G. Bakshi and D.B. Madan, \emph{{Spanning and derivative-security valuation}},
  J. Financ. Econ. 55 (2000), pp. 205--238.

\bibitem{Bates1996a}
D.S. Bates, \emph{{Jumps and Stochastic Volatility: Exchange Rate Processes
  Implicit in Deutsche Mark Options}}, Rev. Financ. Stud. 9 (1996), pp.
  69--107.

\bibitem{Black1973}
F. Black and M. Scholes, \emph{{The Pricing of Options and Corporate
  Liabilities}}, J. Polit. Econ. 81 (1973), p. 637.

\bibitem{Carr1999}
P. Carr and D.B. Madan, \emph{{Option valuation using the fast Fourier
  transform}}, J. Comput. Financ. 3 (1999), pp. 463--520.

\bibitem{Carr2003}
P. Carr and L. Wu, \emph{{The finite moment logstable process and option
  pricing}}, J. Finance 58 (2003), pp. 753--778.

\bibitem{Chourdakis2005}
K. Chourdakis, \emph{{Option Pricing Using The Fractional FFT}}, J. Comput.
  Financ. 7 (2005), pp. 1--23.

\bibitem{Cont2001}
R. Cont, \emph{{Empirical properties of asset returns: stylized facts and
  statistical issues}}, Quant. Financ. 1 (2001), pp. 223--236.

\bibitem{Cont2004}
R. Cont and P. Tankov, \emph{{Financial Modelling with Jump Processes}},
  Chapman {\&} Hall/CRC, 2004.

\bibitem{Crisostomo2014}
R. Crisostomo, \emph{{An Analysis of the Heston Stochastic Volatility Model:
  Implementation and Calibration using Matlab}}, CNMV Work. Pap. No. 58
  (2014).

\bibitem{Duffie2000b}
D. Duffie, J. Pan, and K.J. Singleton, \emph{{Transform Analysis and Asset
  Pricing for Affine Jump-Diffusions}}, Econometrica 68 (2000), pp. 1343--1376.

\bibitem{Fang2009}
F. Fang and C.W. Oosterlee, \emph{{A Novel Pricing Method for European Options
  Based on Fourier-Cosine Series Expansions}}, SIAM J. Sci. Comput. 31 (2009),
  pp. 826--848.

\bibitem{Fiorani2004}
F. Fiorani, \emph{{Option Pricing Under the Variance Gamma Process}}, MPRA Pap.
  No. 15395  (2004).

\bibitem{Gatheral2006}
J. Gatheral, \emph{{The Volatility Surface}}, John Wiley {\&} Sons, Inc., 2006.

\bibitem{Goncu2016}
A. G{\"{o}}nc{\"{u}}, M.O. Karahan, and T.U. Kuzubaş, \emph{{A comparative
  goodness-of-fit analysis of distributions of some L{\'{e}}vy processes and
  Heston model to stock index returns}}, North Am. J. Econ. Financ. 36 (2016),
  pp. 69--83.

\bibitem{Heston1993}
S. Heston, \emph{{A Closed-Form Solution for Options with Stochastic Volatility
  with Applications to Bond and Currency Options}}, Rev. Financ. Stud. 6
  (1993), pp. 327--343.

\bibitem{Hirsa2012}
A. Hirsa, \emph{{Computational Methods in Finance}}, Chapman {\&} Hall/CRC,
  2012.

\bibitem{Itkin2010}
A. Itkin, \emph{{Pricing options with VG model using FFT}}, arXiv Work. Pap.
  (2010).

\bibitem{Jones2003}
C.S. Jones, \emph{{The dynamics of stochastic volatility: Evidence from
  underlying and options markets}}, J. Econom. 116 (2003), pp. 181--224.

\bibitem{Kilin2011}
F. Kilin, \emph{{Accelerating the Calibration of Stochastic Volatility
  Models}}, J. Deriv. 18 (2011), pp. 7--16.

\bibitem{Lee2004}
R.W. Lee, \emph{{Option Pricing by Transform Methods: Extensions, Unification,
  and Error Control}}, J. Comput. Financ. 7 (2004), pp. 51--86.

\bibitem{Lewis2001}
A. Lewis, \emph{{A simple option formula for general jump-diffusion and other
  exponential L{\'{e}}vy processes}}, OptionCity.net Work. Pap.  (2001).

\bibitem{Lewis2013}
A. Lewis, \emph{{Variance Gamma - different results. Willmott Forums}} (2013),
  \urlprefix\url{http://www.wilmott.com/messageview.cfm?catid=8{\&}threadid=95997}.

\bibitem{Lord2010}
R. Lord and C. Kahl, \emph{{Complex logarithms in Heston-like models}}, Math.
  Financ. 20 (2010), pp. 671--694.

\bibitem{Madan1998a}
D.B. Madan, P. Carr, and E.C. Chang, \emph{{The Variance Gamma Process and
  Option Pricing}}, Rev. Financ. 2 (1998), pp. 79--105.

\bibitem{Madan1990a}
D.B. Madan and E. Seneta, \emph{{The Variance Gamma (V.G.) Model for Share
  Market Returns}}, J. Bus. 63 (1990), pp. 511--524.

\bibitem{Matsuda2004}
K. Matsuda, \emph{{Introduction to Option Pricing with Fourier Transform:
  Option Pricing with Exponential L{\'{e}}vy Models}}, Work. Pap.  (2004).

\bibitem{Merton1973}
R.C. Merton, \emph{{Theory of Rational Option Pricing}}, Bell J. Econ. 4
  (1973), pp. 141--183.

\bibitem{Rebonato2004}
R. Rebonato, \emph{{Volatility and correlation: The perfect hedger and the fox:
  Second edition}}, John Wiley {\&} Sons, Inc, 2004.

\bibitem{Ribeiro2004}
C. Ribeiro and N. Webber, \emph{{Valuing path dependent options in the
  variance-gamma model by Monte Carlo with a gamma bridge}}, J. Comput. Financ.
  7 (2004), pp. 81--100.

\bibitem{Ruijter2015}
M.J. Ruijter and C.W. Oosterlee, \emph{{Notes on the BENCHOP implementations
  for the COS method}} (2015).

\bibitem{Schmelzle2010a}
M. Schmelzle, \emph{{Option pricing formulae using Fourier transform: Theory
  and application}}, Prepr. http//pfadintegral.com  (2010).

\bibitem{Schobel1999}
R. Schobel and J. Zhu, \emph{{Stochastic Volatility With an Ornstein-Uhlenbeck
  Process: An Extension}}, Rev. Financ. 3 (1999), pp. 23--46.

\bibitem{Schoutens2004b}
W. Schoutens, E. Simons, and J. Tistaert, \emph{{A Perfect Calibration! Now
  What?}}, Willmott Mag.  (2004), pp. 66--78.

\bibitem{Sepp2003}
A. Sepp, \emph{{Fourier Transform for Option Pricing under Affine
  Jump-Diffusions: An Overview}}, SSRN Electron. J.  (2003).

\bibitem{VonSydow2015}
L. von  Sydow, L. {Josef H{\"{o}}{\"{o}}k}, E. Larsson, E. Lindstr{\"{o}}m, S.
  Milovanovi{\'{c}}, J. Persson, V. Shcherbakov, Y. Shpolyanskiy, S.
  Sir{\'{e}}n, J. Toivanen, J. Wald{\'{e}}n, M. Wiktorsson, J. Levesley, J. Li,
  C.W. Oosterlee, M.J. Ruijter, A. Toropov, and Y. Zhao, \emph{{BENCHOP – The
  BENCHmarking project in option pricing}}, Int. J. Comput. Math. 92 (2015),
  pp. 2361--2379,
  \urlprefix\url{http://dx.doi.org/10.1080/00207160.2015.1072172}.

\bibitem{Zhu2010}
J. Zhu, \emph{{Applications of Fourier Transform to Smile Modeling: Theory and
  Implementation}}, Springer Finance, 2010.

\end{thebibliography}

\end{document}